\documentclass{sig-alternate} 
\usepackage{graphicx,algorithm,algorithmic,epsfig}
\usepackage{amssymb}

\newtheorem{theorem}{Theorem}

\newtheorem{definition}{Definition}
\newtheorem{lemma}{Lemma}

\newtheorem{example}{Example}
\newtheorem{proposition}{Proposition}

\newcommand{\all}{\ensuremath{\, \forall \,}}

\newcommand{\eps}{\varepsilon}

\newcommand{\id}{\textbf{1}}

\newcommand*{\argmin}{\mathop{\mathrm{argmin}}}
\newcommand{\ds}{\displaystyle}

\newcommand{\sta}[1]{#1^{\ast}}

\newcommand{\ac}[1]{\ensuremath{w^{\mathrm{ac}}(#1)}}
\newcommand{\sv}[1]{\ensuremath{w^{\mathrm{sv}}(#1)}}
\newcommand{\SV}{\ensuremath{W^{\mathrm{sv}}}}

\def \gi/{$\gamma$-improve}

\def \apn/{ap-neighbor}

\newcommand{\biwo}{\ensuremath{\mathbf{w}}} 

\newcommand{\trwo}{\mathbf{v}} 
\newcommand{\tilwo}[1]{\ensuremath{\tilde{#1}_S}} 
\def \OPT/{\ensuremath{\mathrm{OPT}}}
 
\newcommand{\wi}[1]{\cup\{#1\}} 
\newcommand{\wo}[1]{\setminus\{#1\}} 
\def\gx/{\mbox{{\tt M$_{\alpha}(\gamma)$}}}
\newcommand{\junk}[1]{}
\newcommand{\rvcg}{\mbox{{\tt REV$_{vcg}$}}}
\newcommand{\rgi}{\mbox{{\tt REV$_{\gamma,\alpha}$}}}

\junk{
\title{Online Ad Slotting With Cancellations}
\numberofauthors{2}
\author{
\alignauthor
\mbox{Florin Constantin}\titlenote{Most of this work was done while the author was visiting Google Research, New York.}\\
\affaddr{SEAS, Harvard University} \\
       \affaddr{33 Oxford St.}\\
       \affaddr{Cambridge, MA 02138}\\
       \email{florin@eecs.harvard.edu}
\alignauthor
\vspace{-0.12in} 
\mbox{Jon Feldman, S. Muthukrishnan and Martin P\'al}
\\
       \affaddr{Google Research, New York}\\
       \affaddr{76 9th Avenue, 4th floor}\\
       \affaddr{New York, NY 10011}\\
       \email{\{jonfeld,muthu,mpal\}@google.com}
}
}

\newcommand{\aut}{\raisebox{8pt}{$\dagger$}} 
\title{Online Ad Slotting With Cancellations}
\numberofauthors{4}
\author{
Florin Constantin\titlenote{
SEAS, Harvard University, 33 Oxford St., Cambridge, MA 02138. Email: {\tt florin@eecs.harvard.edu}.
Most of this work was done while the author was visiting Google Research, New York.
}
\alignauthor
Jon Feldman\titlenote{
Google Research, 76 Ninth Ave., 4th Floor, New York, NY, 10011. Email: {\tt \{jonfeld,muthu,mpal\}@google.com}.
}
\alignauthor
S. Muthukrishnan\aut
\alignauthor
Martin P\'al\aut
}

\begin{document}
\maketitle 

\begin{abstract}
Many advertisers use Internet systems to buy advertisements 
 on publishers' webpages or on traditional media such as radio, TV and 
newsprint. They seek a simple, online mechanism to {\em reserve} ad slots in advance. 
On the other hand, media publishers represent a vast and varying inventory, and 
they too seek automatic, online mechanisms for pricing and allocating such reservations.

In this paper, we present and study a simple model for auctioning such
ad slots in advance.  Bidders arrive sequentially and report which
slots they are interested in.  The seller must decide immediately whether or not to grant a reservation.  Our model allows a
seller to accept reservations, but possibly {\em cancel} the
allocations later and pay the bidder a cancellation compensation ({\em
bump payment}).

Our main result is an online mechanism to derive prices and 
bump payments that is efficient to implement. This mechanism has many desirable properties. 
It is individually rational; winners have an incentive to be honest and bidding one's true value dominates any lower bid. 
 Our mechanism's efficiency is within a constant fraction of the {\em a posteriori} optimally efficient solution. Its revenue is within a constant fraction of the {\em a posteriori} VCG revenue. 
Our results make no assumptions about the order of arrival of bids or
the value distribution of bidders.  All our results still hold if the items for sale are elements of a matroid, a more general setting than slot allocation.
%
\end{abstract}

\section{Introduction}
Many advertisers now use Internet advertising systems. These take the
form of advertisement ({\em ad}, henceforth) placements either in
response to users' web search queries, or at predetermined slots on
publishers' web pages.
In addition, increasingly, advertisers use Internet systems that sell ad slots on behalf of
offline publishers on TV, radio or newsprint.
In sponsored search, and in some other cases,
ad slots are typically sold via 
{\em real time} auctions, i.e., when a user poses a query or visits a web page, an auction is used to determine which ads will show and where 
they will be placed. On the other hand, traditionally, advertisers  seek
ad slots {\em in advance}, i.e. to reserve their slots. Product releases (such as movies, electronic gadgets, etc) and ad campaigns
(e.g., creating and testing ads, budgets) are planned ahead of time and need to coordinate with future events
that target suitable demographics. The advertisers then do not want to risk the vagrancies of real-time auctions
and lose ad slots at critical events; they typically like a reasonable guarantee of ad slots at a specific time in
the future within their budget constraints today.

Our motivation arises from systems that enable such advanced ad slotting. 
In particular, our focus is on automatic systems that have to manage ad slots in many different publishers' properties. 
These properties differ wildly in their traffic, targeting, price and effectiveness;
consider placement in the front page of {\tt nytimes.com} versus placement in an individual's blog versus a radio slot in a local country
music radio station. Also, the inventory levels are massive. Slots and impressions in web publishers' 
properties as well ad slots in TV, radio, newsprint and other traditional media are in 100's of
millions and more. Not all publishers can estimate their inventory accurately: traffic to websites responds to 
time-dependent events, and sometimes webpages are generated dynamically so that even the availability of a slot in the future is
not known {\em a priori}.
Most  web publishers 
are not able to estimate accurately a price for an ad slot, or provide sales agents to negotiate terms and would like automatic 
methods to price ad slots. Thus, what is desirable is a simple, automatic, online\footnote{We use the
word {\em online} as in {\em online algorithm}---i.e., the input
arrives over time, and the algorithm makes sequential decisions ---we
do not mean ``on the Internet.''} market-based mechanism to enable advanced ad slotting
over such varied, massive inventory. 

Inspired by these considerations, we study the problem of mechanism design for advanced ad slotting. The problem is quite general with many
facets. Our contribution is to propose a simple model, to design a suitable mechanism and to analyze its properties.
In more detail, our contributions are as follows.

\medskip
\noindent
{\bf (i)} We propose the following simple high-level model for
advanced ad slotting auctions.  An auction starts at time $0$; the
seller has a set of {\em slots} for sale that will be published at
time $T$.  Bidder $i$ arrives at some time $a_i < T$, reports which
slots he is interested in, places a bid $w(i)$ and requests an immediate response.  Bidder $i$ is either accepted or rejected; if accepted, he may
be removed ({\em bumped})  later, but in that case, he is
awarded a {\em bump payment}.  We assume that if bumped, a bidder incurs a loss of an $\alpha$ fraction of its value.
At time $T$, each accepted bidder that
has not been bumped is published in one of the slots he was interested
in, and pays a price that is at most his bid.

This model lets the publisher accept a reservation at time $t$ for a slot 
available at a later time $T$, and lets the advertiser get a
reasonable guarantee. However, crucially, it lets the publisher {\em
cancel} the reservation at a later time. Cancellation is necessary for
publishers to take advantage of a spike in demand and rising prices
for an item and not be forced to sell the slot below the market
because of an {\em a priori} contract.  In addition, in a pragmatic
sense, cancellation is crucial: for example, a website might
overestimate its inventory for a later date and accept ads, but as
time progresses, its estimates may become smaller, and the publisher
will not be able to honor all the accepted ads from the past. Finally,
cancellations are very much part of the business with advance
bookings, both within advertising and beyond. At the same time, it
comes at a cost, which is the bump payment. This is reasonable since
it compensates the advertiser for the uncertainty, and lets
advertisers recoup part of their costs involved in preparing a
campaign for time $T$ based on their reservation at time $t$. We present our model formally in Section~\ref{sec:model}.


\medskip
\noindent
{\bf (ii)} We present an efficiently implementable mechanism \gx/ for determining who is accepted, who is bumped and also the prices and bump payments. The parameter $\gamma$ represents how much higher a new bid has to be in order to bump an older bid.
 A bumped bidder will be paid an $\alpha$ fraction of their bid, making up for their utility loss due to being bumped.

\medskip
\noindent
{\bf (iii)}
We show a number of important strategic as well as
efficiency- and revenue-related properties of \gx/: 
\begin{itemize}
\item
 \gx/ is individually rational and winners have an incentive to bid truthfully while losers 
should bid at least their true value. 

\item
 With respect to the bids received, the efficiency (value of assignment) of \gx/ is  at least a constant factor (depending on $\gamma$ and $\alpha$) of the offline optimum. 
Under mild player rationality assumptions, we show that 
our mechanism is competitive with respect to the optimum offline efficiency on bidders' true {\em values}. 

\item
We also consider the notion of {\em effective efficiency} which interprets social welfare as the
sum of the winners' bids  minus  bumped  (if any) bidders' losses.  We show that for
suitable $\gamma(\alpha)$, our mechanism's effective
efficiency matches a numerically obtained upper bound on the effective efficiency
of {\em any} deterministic algorithm. To our knowledge, costly cancellations have not been previously studied in online bipartite matching problems.

\item
 The revenue of \gx/  is at least a constant factor (dependent on $\gamma$ and $\alpha$) of that of the 
Vickrey-Clarke-Groves (VCG) mechanism on all received bids. 

\item
We also study {\em speculators}, that is, ones who have no interest in the items for sale but who participate in order to earn the 
bump payment. We show several game theoretic properties about the behavior of the speculators, including bounding their overall profit. 
\end{itemize}

To the best of our knowledge our results are the first about
mechanisms with strong game-theoretic properties for advanced
placement of ads (more generally, indivisible goods) with a
cancellation feature. We make no assumptions on the arrival order of
the bidders or on their values.  Prior work has studied advanced
sale of goods without cancellations, but only under a probabilistic
distribution of bidders'
values~\cite{GallienGupta,lavi00competitive}. Under a worst case model
like ours, no nontrivial results are possible without making
additional assumptions; in our case, we overcome these impossibilities
by allowing cancellations. 
 In secretary problems~\cite{matroid}, bids may be
arbitrary but their order is assumed to be uniformly random (cannot be specified by
an adversary).


There are specific examples of systems that implement advanced booking with cancellations. For example, 
this is common in the airline industry, where tickets may be booked ahead of time, and 
customers may be bumped later for a payment. In the airline case, the inventory is mostly fixed, sophisticated models are used to calculate prices over time,
and often negotiations are involved in establishing the payment for bumping, just prior to time $T$. In some cases, the bump payments may even be larger than the original bid (price) of the customer. Likewise, in offline media such as TV or Radio,
humans are involved are in negotiating advanced prices, and often if the publisher does not respect the reservation due to inventory crunch,
a payment is {\em a posteriori} arranged including possibly a better ad slot in the future. These methods are not
immediately applicable to the auction-driven automatic settings like ours. 

From a technical point of view, one can view our model as an online
weighted bipartite matching problem (or more generally, an online
maximum weighted independent set problem in a matroid).  On one side
we have slots known ahead of time.  The other side comprises
advertisers whose bids (weighted nodes) arrive online. Our goal is to
find a ``good'' weighted matching in the eventual graph. Each time an
advertiser appears we need to decide if we should retain it or discard
it; retaining it may lead to discarding a previously retained
bidder. Our mechanism builds on such an online matching
algorithm~\cite{McGregor} to determine a suitable bump payment and
prices. It is, curiously, able to make use of such an online algorithm
previously proposed in the {\em semi-streaming} model in the
theoretical computer science literature.

All our results extend to a setting where the items for sale are elements of a matroid, a more general setting than slot allocation. A bidder bids on exactly one element of the matroid, which is known ahead of time and may vary across bidders.
A set $S$ of bidders is then feasible if the set containing each bidder's element  forms an independent set of the matroid. In the bipartite matching setting, the seller's matroid contains one element for each subset of slots and a set of bidders (elements) is independent if the bidders can be matched to slots such that each one receives an element from its subset. 
We prefer the matching language for clarity of exposition. 

We have initiated the study of mechanisms for advanced reservations
with cancellations. A number of technical problems remain open, both
within our model, as well as in its extensions, which we
describe later for future study.

\section{Related Work}

There is considerable work on auctions when bidders are present throughout 
a period of time. Of more direct relevance are the following classes of problems.

Babaioff et al.~\cite{matroid} address the {\em matroid secretary problem}: finding a competitive assignment when weighted elements of a matroid arrive online and no cancellations are allowed.
As is common in secretary problems, while not making any assumption on bidder valuations, they assume that all orders of arrivals are equally likely. They present a $\log r$-competitive algorithm for general matroids where $r$ is the rank of the matroid (the size of the largest independent sets) and a $4d$-competitive algorithm for our setting without cancellations (transversal matroids) but where each bidder can only be interested in at most $d$ items. 
Both these algorithms observe half of the input and then set a threshold price:  per item in the transversal case and uniform in the general case. 
 The $4d$-competitive algorithm ensures truthful bidding even when the items desired by an agent are private information. 
Dimitrov and Plaxton~\cite{texasMatr} extend the $4d$-algorithm and provide an algorithm with a  constant competitive ratio for any transversal matroid.

Bikhchandani et al.~\cite{Vohra} present an ascending auction for selling elements of a matroid that ends with an optimal allocation (i.e. the auction is efficient). Truthful bidding is an equilibrium of the auction. They assume however that bidders are present throughout the auction.


Cary et al.~\cite{flaxman} show that a random sampling profit extraction mechanism approximates a VCG-based target profit in a procurement setting on a matroid.

Gallien and Gupta~\cite{GallienGupta} analyze players' strategies regarding buyout prices in online auctions. 
In their model, bidders' valuations are drawn from a known distribution and their utilities are time-discounted;
furthermore there are no cancellations and arrivals are assumed to follow a Poisson process. 
They exhibit symmetric Bayes-Nash equilibria in which buyers follow certain threshold strategies.
 Assuming that buyers follow the corresponding equilibrium strategies, the seller can then optimize revenue by tuning the price function.

Lavi and Nisan~\cite{lavi00competitive} consider online auctions for identical goods. In their model, bidders' values are arbitrary from the interval $[\underline{\rho},\overline{\rho}]$ and no cancellations are possible. They present a simple online posted-price auction based on exponential scaling. This auction is optimal among online auctions and achieves a $\Theta(\log(\overline{\rho}/\underline{\rho}))$ approximation with respect to both efficiency and the VCG revenue.

Independently form us, Feige et al.~\cite{bannerPenalties} study an {\em offline} weighted bipartite matching problem where the seller can partially satisfy a bidder's request at the cost of paying a proportional penalty. Accepting a bid but not providing any items results in a utility loss proportional to the bid, similar to our definition of {\em effective efficiency} in Section~\ref{subsec:effeff}. They show that it is NP-hard to approximate the optimal solution within any constant factor. They propose an adaptive greedy algorithm assigning one bidder at a time (but that inspects all unassigned bidders in deciding which bidder to allocate) and that may reassign bidders. They provide a lower bound on this algorithm's efficiency with respect to the optimal assignment. 

\section{Model and Mechanism}
\label{sec:model}

\newcommand{\idac}{\ensuremath{\id_\mathrm{ac}}}
\newcommand{\CI}{\mathcal{I}}

We first define our model and present our mechanism which we will
study in later sections. 

\subsection{Model Basics}

There is a seller who has a finite set of {\em slots} (items), and starts sale at time $0$ and 
ends sale at time $T$. 

Each bidder $i$ is interested in exactly one slot out of a set $N(i)$ called $i$'s {\em choice set}. 
We denote by $v(i)$ bidder $i$'s value for any slot in $N(i)$ and we assume that $v(i)$ is private information to $i$. 
Each bidder $i$ places a bid $w(i)$ (that may be different than $v(i)$)  as soon as it arrives, at time $a_i$.
As a consequence of $i$ bidding, $i$'s choice set becomes known to the seller.
%
When $i$ bids, he may be {\em accepted} (i.e. promised an item from $N(i)$), else rejected. If promised an allocation,
he may get {\em bumped} later, losing the reservation. Any accepted bidder who is not bumped before time $T$ is allocated.
We model bidder $i$'s {\em utility} as 
\begin{equation}
\label{eq:ut}
\lambda \cdot v(i) - x(i), \quad \mbox{where}
\end{equation}
\begin{itemize}
\item $\lambda$ equals $0$ if $i$ is rejected, $1$ if $i$ is accepted and granted from $N(i)$ and $-\alpha$ if $i$ is accepted but bumped.
\item $x(i)$ is $i$'s {\em transfer} to the seller (price). It is $0$ if $i$ is rejected,
and some non-negative amount if $i$ is accepted and allocated. 
 $x(i)$ may be negative (e.g. the {\em bump payment} the seller makes in $\gx/$).
%

\end{itemize}

That is, a bidder is unaffected if rejected right away, has a value of $v(i)$ for being allocated, and incurs a loss amounting to an $\alpha$ fraction of its value if bumped. The utility is quasilinear in money.

Note that from an algorithmic perspective, our problem is online maximum weighted bipartite matching with costly cancellations. There is a bipartite graph with items (respectively bidders) as ``right''-(respectively ``left''-)hand side vertices. Left-hand side vertices are fixed. One by one, a right-hand side vertex $i$ is revealed together with its weight $w(i)$ and its edges (i.e. the set $N(i)$). A decision whether to accept $i$ or not must be taken immediately.
The goal is to find a matching of weight as high as possible, where cancellations are allowed, but canceling bidder $i$ of weight $w(i)$ results in a {\em penalty} of $\alpha w(i)$.

However, in our setting bidders are self-interested and may alter the input to the algorithm (their bids) if it is in their interest to do so. Therefore we aim for a mechanism that is competitive while  bounding the bidders' manipulations. 
Note that if $\alpha = 0$, then we can tentatively accept any bidder
and only decide at time $T$ which bidders are truly accepted while
rejecting the remainder, thus reducing the problem to finding a bipartite matching of maximum weight, 
 a standard offline optimization. By charging each bidder its VCG (see Section~\ref{sec:rev}) price, it becomes a dominant strategy for each bidder to be {\em truthful} (i.e. bid its true value).



\subsection{Our $\gx/$ Mechanism}

We present our advance-booking online mechanism $\gx/$ (allocation algorithm and payments). 
 The allocation algorithm follows the {\it
Find-Weighted-Matching} algorithm in~\cite{McGregor}\footnote{Unlike~\cite{McGregor},
a bidder $i$'s value is the same for any slot (vertices as
opposed to edges are weighted). Our mechanism may then change the slot
$i$ is currently assigned to at various stages in the
algorithm.}. Apart from $\alpha$, which is specified by the model, the
algorithm uses an {\em improvement factor} $\gamma>0$ such that $0 < \alpha <
\frac{\gamma}{1+\gamma}$.
  
Denote the number of bidders by $n$; our mechanism is independent of $n$.
By relabeling bidders, assume that they bid in order $1,2, \dots, n$; time is
indexed likewise. 

\begin{definition}
We say that a set of bidders $B$ {\em can be matched} if for each $b \in B$ there exists an item $i_b \in N(b)$ such that $i_b \neq i_{b'}, \all b \neq b'$. We say that $B$ is a {\em perfect matching} if it can be matched and no item is left unmatched ($B$'s cardinality must equal the number of items).
\end{definition}

\gx/'s pseudocode is listed in Algorithm~\ref{algo:gi}.  At a high
level, \gx/ maintains a set of accepted bidders that form a perfect matching.
For any new
arriving bidder $i$ bidding $w(i)$, the algorithm looks if there exists some bidder $j$ in the accepted set with
$w(j) < \frac{w(i)}{1+\gamma}$ such that $i$ can be swapped in the matching if $j$ is 
swapped out. If so, 
 accept $i$ and cancel the reservation of ({\em bump}) $\sta{j}$, the lowest weight such $j$.  Bidder $\sta{j}$ is paid the bump payment $\alpha w(\sta{j})$. (Note
that $\sta{j}$ makes no payment at all and in fact {\em gets money for free} from the
seller if bumped.) 
An accepted bidder who is not bumped by time $T$ is necessarily allocated a slot from his choice set and pays the seller an amount we define later (Eq.\eqref{eq:p}).

At time 0, $A_0$ is an arbitrary matching;  
we introduce $r$ dummy bidders (each bidding 0) whose choice set is the whole set of items, arriving before all actual bidders. 
This will not affect our arguments below.\footnote{
When bidder $t$ arrives, 
assume $A_{t-1} = A \cup D$ where $D$ only contains dummy bidders and there exists a matching $I_t$ of $A\wi{t}$ which matches $t$ to some item $i_t$. 
By reassigning dummy bidders, we can assume that actual bidders are matched according to $I_t$. 
Then bidder $t$ can bump at least the dummy bidder $d\in D$ that is matched to $i_t$ in $A_{t-1}$.
}
At time $t$, we call currently accepted bidders {\em alive}, and denote the set of 
alive bidders as $A_t$. Let $X_t = \{b \in A_{t-1}: A_{t-1} \wi{t} \wo{b}$ can be matched$\}$;
$X_t$ is the set of alive bidders at $t-1$ that can be exchanged for $t$. 
Bidders still alive at the end (time $T$) are called {\em survivors} and 
$S=S(\biwo)$ denotes this set. 
We denote the set of bumped bidders by $R=R(\biwo)$. 

%
\begin{algorithm}[th]
\begin{algorithmic}
\STATE Fix $A_0$, an arbitrary perfect matching on dummy bidders.
\FOR{each $i\geq 1$ bidding $w(i)$ (with choice set $N(i)$)} 
\STATE Let $X_i=\{j<i: A_{i-1} \wi{i} \wo{j} $ can be matched$\}$.
\STATE Let $\sta{j} = \argmin_{j \in X_i} w(j)$
\IF{$(1+\gamma)w(\sta{j}) < w(i)$}
\STATE  $ A_i = A_{i-1} \wi{i} \wo{\sta{j}}$:  $i$ is accepted, $\sta{j}$ is bumped  
\STATE $\sta{j}$ is paid $\alpha w(\sta{j})$, i.e. an $\alpha$ fraction of its bid
\ENDIF
\ENDFOR
\STATE Charge any bidder $i$ in $S \triangleq A_n$ (i.e. survivor) 
 as in Eq.~\eqref{eq:p}.
\end{algorithmic}
\caption{\gx/.}
\label{algo:gi}
\end{algorithm}


\begin{definition}
Let $i$ be a bidder and fix the bids of all other bidders. 
Let \ac{i} ($i$'s {\em acceptance weight}) be the infimum of all bids that $i$ can make such that $i$ is accepted when he bids. Similarly, 
let \sv{i} ($i$'s {\em survival weight}) be the infimum of all bids that $i$ can make such that $i$ is accepted when he bids
{\em and survives until time $T$ (the end)}. Clearly, $\ac{i} \leq \sv{i}$.  Let $\SV{} = \sum_{i \in S} \sv{i}$.
\end{definition}

Note that $\sv{i}$ always exists since it suffices to bid
$(1+\gamma)\max_{j\neq i} w(j)$.  Also, $\ac{i}$ and $\sv{i}$ are
independent of $i$'s actual bid, but may depend on the time $i$ bids
and on the other bidders' bids or arrivals (times of
bidding). Note that $\ac{i}$ can be computed by the seller as soon as a bidder
arrives whereas $\sv{i}$ may depend on future bidders and thus can
only be computed at time $T$.



In summary, $i$ is 
$
\begin{cases}
\mbox{rejected}, & \text{ if $w(i)<\ac{i}$} \\
\mbox{bumped}, & \text{ if $\ac{i} \leq w(i)<\sv{i}$} \\
\mbox{a survivor}, & \text{ if $\sv{i} \leq w(i)$} 
\end{cases}
$
If $i$ is a survivor, $i$'s price $p_i$ is as follows:
\begin{equation}
\text{\hspace{-0.1in}}
\label{eq:p}
p_i =
\begin{cases}
 \sv{i} (1 - \alpha) & 
\text{if $\ac{i}<\sv{i}$.}
\\
 \sv{i} &  \text{\hspace{-0.1in}if } \ac{i}=\sv{i}. 
\end{cases}
\end{equation}

The common case is when $\ac{i} < \sv{i}$: $i$ gets a discount amounting to the highest refund it could have otherwise obtained: $\alpha \sv{i}$. 
The special case of $\ac{i} = \sv{i}$ occurs when $i$'s acceptance is enough for its survival (in particular if $i$ is the last bidder). 
When $\ac{i} = \sv{i}$, from the bidder point's of view, \gx/ posts a price of $\ac{i}$.


This concludes the definition of \gx/. We will now  study its properties.


\section{Incentive Properties} 
%

The following theorem summarizes our mechanism's favorable incentive properties. 
\begin{theorem}
\label{thm:incentives}
The \gx/ mechanism is individually rational.  Bidding  one's true value (weakly) dominates any lower bid. If honest, any survivor is (weakly) best-responding. 
\end{theorem}
The proof follows from the two lemmas below.

Note that with bump payments (``money for nothing'') we cannot hope to have a truthful mechanism since anyone with no interest in any allocation can bid hoping to get a bump payment. 
 It is nevertheless possible that other types of truthful competitive allocation mechanisms exist. 

\begin{lemma}
\label{prop:BR}
%
Bidding less than one's true value is dominated by bidding one's true value.  A bumped bidder's best response 
may however be to bid more than its true value.
\end{lemma}
\begin{proof}
If $\ac{i}<\sv{i}$, 
bidder $i$'s highest possible bump payment is $\alpha \sv{i}$. The price of $(1-\alpha)\sv{i}$ has been 
chosen such that $i$ prefers winning to being paid $\alpha \sv{i}$  if and only if $v(i) \geq \sv{i}$. That is, the best bid $i$ can make ($i$'s {\em best-response}) is to bid just below $\sv{i}$ if $v(i) < \sv{i}$ and to bid its true value otherwise.

If $\ac{i}=\sv{i}$, then $i$ can never get a bump payment and $i$ simply faces a take-it-or-leave-it offer of $\sv{i}$.
\end{proof}


Can a bidder ever incur a loss by participating? The following result shows that if truthful, no bidder will have a negative utility, i.e. the mechanism is individually rational.
\begin{lemma}
\label{lemma:IR}
If bidder $i$ bids its true value $v(i)$, 
 then $i$'s utility after participating in the mechanism is non-negative.
\end{lemma}
\begin{proof}
When surviving, $i$ pays at most $\sv{i} \leq v(i) $. If $i$ is not accepted then $i$'s utility is 0. If $i$ is accepted and then bumped, $i$'s utility is $-\alpha v(i) + \alpha v(i) = 0$.
\end{proof}
%


In the following two sections we show that apart from favorable incentive properties,  \gx/ is also competitive with respect to revenue and efficiency.

\section{Efficiency of \gx/}
\label{sec:eff}

\begin{definition}
For any
vector $\biwo'=(w'(1), \dots, w'(n))$ of weights, we let $\OPT/[\biwo']$ be the weight of the optimal matching (recall that the seller knows any bidder's choice set).
We will overload the expression and let $\OPT/[\biwo']$ also denote the 
optimal matching (as opposed to its weight), when there is no confusion.
If $B$ is a set of bidders, we will denote  $w'(B) = \sum_{b \in B} w'(b)$. 
Unless specified otherwise, $\biwo
=(w(1), \dots, w(n))$ denotes the input bids and $\trwo$ denotes the
true values. We will denote by $\OPT/ = \OPT/[\biwo]$. 
\end{definition}

From an algorithmic perspective, our mechanism is a $1+\gamma$ approximation to the optimum assignment (Lemma~\ref{result:w(S)}). However, if incentives are not aligned then bidders may want to  significantly alter the input to the algorithm (their bids). The allocation may then be a poor choice considering bidders' {\em true} values. 
%
We show that this is not the case, in 
 Theorem~\ref{prop:effTruePos},  our main result regarding \gx/'s efficiency: 
%
  the assignment output by  our mechanism is  a constant factor (depending on $\alpha$ and $\gamma$) approximation to the offline optimum on bidders' {\em true values} if bids are "reasonable".
%
\begin{theorem}
\label{prop:effTruePos}
Let $\biwo$ be a set of bids such that each bidder bids at least its true value,
that is $w(i) \geq v(i) \all i$, and the sum of all bidders' utilities is non-negative. 
When run on $\biwo$, \gx/'s  efficiency 
with respect to  the {\em true values} $\trwo$ is:
\[
\sum_{i \in S(\biwo)} v(i) \geq \frac{1 - \alpha - \frac{\alpha}{\gamma} }{(2 - \alpha - \frac{\alpha}{\gamma})(1+\gamma)} \cdot \OPT/[\trwo].
\]
\end{theorem}
Note that if all bidders are truthful then the right-hand side constant can be increased to $\frac{1}{1+\gamma}$ (see Lemma~\ref{result:w(S)} below).

Recall the impossibility of making truthfulness a dominant strategy due to the use of bump payments. Theorem~\ref{prop:effTruePos}'s assumption allows for some bidders to have negative utility and therefore fail to best-respond (Lemma~\ref{lemma:IR} shows that simply by being truthful, a bidder's utility is at least 0) as long as overall, gains outweigh losses in utility. The assumption fails when, for example, bidders with value 0 for an allocation (the ``speculators'' of Section~\ref{sec:spec}) grossly overestimate the actual bids and end up being allocated and having to pay due to bidding too high. In such a scenario, the true value of the allocation may be very small, possibly 0. We refer the reader to Section~\ref{sec:spec} for a more detailed discussion on how speculators affect incentives.

We now proceed to proving Theorem~\ref{prop:effTruePos}, establishing a few other important results along the way. The following bid vector will prove useful: 
$$
\tilwo{\biwo}(i)=
\begin{cases}
\sv{i}, & \mbox{ if $i\in S$} \\
w(i)/(1+\gamma), & \mbox{ if $i\notin S$} 
\end{cases}.
$$
 Note that  if $i\notin S$ then $\sv{i} > w(i) > w(i)/(1+\gamma)$.

 Lemma~\ref{lemma:funnyvcg} is used for both efficiency and revenue claims. Its proof is more involved and we defer it to the Appendix.
\begin{lemma}
\label{lemma:funnyvcg}
 Any survivor $ s \in S(\biwo)$ is also in
$\OPT/[\tilwo{\biwo}]$.
\end{lemma}
%
%

The following result provides an upper bound on the sum of bumped bidders' bids and shows that $S$ is a $1+\gamma$ approximation to the
optimal offline matching given the same set of bids.  Recall that $R$ is the set of bumped bidders.
\begin{lemma}
\label{result:w(S)}
We have
$w(R) \leq $
$\SV{}/\gamma \leq w(S)/\gamma$.
Also, $\OPT/ \leq (1+\gamma) w(S)$.
\end{lemma}
\begin{proof}
For each $s\in S$, let $d_1, \ldots, d_J = s$ be a chain such that: $d_{j+1}$ bumps $d_j$, $\all 1\leq j \leq J-1$. To simplify notation, assume $d_1=1, \dots, d_{J-1}=J-1$. We will show that
\[
\sum_{j=1}^{J-1} w(j) \leq \sv{s} / \gamma
\]
The claim will follow since the set of bidders is the disjoint union of survivors' chains.

Since $s$ bumped $J-1$, we have $w_{J-1} \leq \frac{\sv{s}}{1+\gamma}$.
Since $j+1$ bumps $j$, $\all 1\leq j \leq J-2$, $ w_j \leq \frac{w_{j+1}}{1+\gamma}$. 
Thus by induction, $w_j \leq \sv{s} (1+\gamma)^{j-J}, \all 1\leq j \leq J-1$. We get 
\[
\sum_{j=1}^{J-1} w_j \leq  \sv{s} \sum_{j=1}^{J-1} (1+\gamma)^{j-J} 
\leq \sv{s} / \gamma
\]

We have $w(S) \geq \tilwo{\biwo}(S) = \OPT/[\tilwo{\biwo}] \geq
\OPT/[\biwo]/(1+\gamma)$. Each inequality is implied by the fact that
no bidder's contribution decreases when going from the left hand side
to the right hand side. The equality follows from
Lemma~\ref{lemma:funnyvcg}: $S$ is
an optimal basis for $\tilwo{\biwo}$, i.e. $\tilwo{\biwo}(S) =
\OPT/[\tilwo{\biwo}]$.
\end{proof}
An analogous lemma can be found in~\cite{McGregor}. Our constants are
tighter because in our model, a
bidder's value for any slot is the same,  and all edges incident
to a bidder arrive simultaneously. These bounds are almost tight:
\begin{example}
\label{ex:tightk+2}
Consider $k+2$ truthful bidders competing on one item; bidder $i$ is the $i$-th to arrive and has value $(1+\gamma)^{i-1}$ unless $i=k+2$, whose value is $(1+\gamma)^{k+1}-\eps$. 
Bidder $i+1$ bumps $i$, $\all 1 \leq i \leq k$.  Only the $k+1$-st bidder survives. 
The bumped bidders have total weight $\sum_{i=0}^{k-1} (1+\gamma)^{i} = ((1+\gamma)^k - 1)/\gamma$. 
\OPT/ is $(1+\gamma)^{k+1}-\eps$.
\end{example}
\begin{proof}[of Theorem~\ref{prop:effTruePos}]
  By assumption, bidders' total utility is non-negative:
\begin{align*}
\sum_{s\in S} \left( v(s) - (1-\alpha) \sv{s} \right) &+ \sum_{r \in R} \alpha w(r) \geq 0 
\end{align*}
We have from Lemma~\ref{result:w(S)} that 
$\sum_{r \in R} w(r) \leq \sum_{s\in S} \sv{s} /\gamma$.  
We prove that 
$$\sum_{s\in S} \left(v(s) + \sv{s}\right) \geq \OPT/[\trwo] / (1+\gamma).$$
 The theorem then follows by algebraic manipulation.

Let $w'(s) := \max(v(s), \sv{s}) \all s\in S$. 
Clearly, $w(s) \ge w'(s) \all s\in S$, otherwise they would not have survived. Also, $v(s) + \sv{s} \ge w'(s) \all s\in S$ by definition.
If everyone in $S$ bids $w'$ instead of $w$, the outcome does not change (they still bid above their survival thresholds). By Lemma~\ref{result:w(S)}, 
$\sum_{s\in S} w'(s) \ge \OPT/[\biwo'] / (1+\gamma)$ and $\OPT/[\biwo'] \ge \OPT/[\trwo]$ by definition, giving the claim.
\end{proof}


\subsection{Effective Efficiency}
\label{subsec:effeff}
%

Efficiency is usually measured as the sum of winning bidders' values. 
An alternative definition which also takes into account the losses of bumped bidders and may be more appropriate when cancellations are allowed is the following:

\newcommand{\compRatio}{\ensuremath{c}}

\begin{definition}
Let $w$ be a sequence of bids and $A$ an online allocation algorithm, possibly with cancellations. Let  $S(A)$ (resp. $R(A)$) be the set of winners (resp. bumped bidders) when $A$ is run on $w$. 
 We define the {\em effective efficiency} of $A$ on $w$ as 
\[ \textstyle
u(w) = \sum_{s \in S(A)} w(s) - \alpha \sum_{r \in R(A)} w(r)
\]
$A$'s  effective efficiency {\em competitive ratio} is
$ \ds
\inf_w \frac{u(w)}{\OPT/[w]}
$
\end{definition}

We present  an upper bound (obtained numerically) on the competitive ratio of {\em any} deterministic algorithm.
 For $\alpha < 0.618$ and a certain $\gamma_\alpha$, $M_\alpha(\gamma_\alpha)$ matches this upper bound.

For a fixed $\alpha$, let $n \geq 2$ a positive integer and $\compRatio \in (0,1)$: we aim for $n$ bidders and a competitive ratio of $\compRatio$. 
Consider one item and a sequence of bids $\{a_k (\compRatio)\}_{1\leq k \leq n}$ on it (bidder $k$ bids $a_k$) such that \ %
$a_1 = 1, \ a_2 \ds = \frac{1}{\compRatio} > 1$ \text{ and} 
\begin{align}
\compRatio a_{k+1}(\compRatio)  = a_k(\compRatio)  - \alpha \sum_{j=1}^{k-1} a_j(\compRatio) \all k \geq 2 \text{ implying}\notag
\\
\compRatio a_{k+1} = (1+\compRatio) a_k - (1+\alpha) a_{k-1} \all k \geq 2 \label{eq:rec}
\end{align}
We will look for a $\compRatio = \compRatio_n$ such that
\begin{equation}
  \label{eq:la}
\text{\hspace{-0.25in}}  a_n(\compRatio) - \alpha \sum_{j=1}^{n-1} a_j(\compRatio) = \compRatio a_n(\compRatio) \iff a_n = (1+\alpha) a_{n-1}
\end{equation}
$ \ds \text{E.g. }
\compRatio_2 = \frac{1}{1+\alpha} > \compRatio_3 = \frac{1}{1+2\alpha} > \compRatio_4 = {\textstyle \frac{2}{1+3\alpha + \sqrt{(1+5\alpha)(1+\alpha)}}.}
$
Unfortunately, $\compRatio_n$ does not have a nice closed form for $n \geq 4$ (in addition, $\compRatio_n$ may be not be unique - the smallest $\compRatio_n \in [0,1]$ is then of interest).

%
\begin{theorem}
\label{prop:effEff}
Fix $n$ and $\alpha$. Let $\compRatio_n$ be the lowest number (if any) in  $[0,1]$ for which Eqs.~\eqref{eq:rec} and~\eqref{eq:la} simultaneously hold. Then no deterministic algorithm can have an effective efficiency  competitive ratio  higher than $\compRatio_n$.
\end{theorem}
\begin{proof}
On any input, the offline optimum with respect to effective efficiency is simply the highest weight assignment, and it results in bumping no bidders. 

Assume that the bids that arrive are $a_1, \dots, a_{k_0}$ for some $1 \leq k_0 \leq n$. Then 
at each $k$, the algorithm $A$ must accept $a_k$, or its competitive ratio will be smaller than $\compRatio_n$ when $k=k_0$. This is clear for $k=1$. 
Fix  $k \in [2,n-1]$. Let $M_k$ be the highest (i.e. the offline optimum) of $a_1, \dots, a_k$.
 If $A$ does not accept $k$ then the competitive ratio on input $a_1, \dots, a_k$ will be at most
\[
\frac{a_{k-1}(\compRatio_n)  - \alpha \sum_{j=1}^{k-2} a_j(\compRatio_n)}{M_k(\compRatio_n)}
=
\frac{\compRatio_n a_k(\compRatio_n)}{M_k(\compRatio_n)} 
\leq
\compRatio_n 
\]
where the equality follows from Eq.~\eqref{eq:rec}.
Now we claim that {\em whether or not} $A$ accepts $a_n$, its competitive ratio will be at most $\compRatio_n$. If $a_n$ is accepted,  $\alpha \sum_{j=1}^{n-1} a_j$ has been lost due to bumping bidders $1, \dots, n-1$; if $a_n$ is rejected the effective efficiency is  $a_{n-1} - \alpha \sum_{j=1}^{n-2} a_j$. By Eqs.~\eqref{eq:rec} and~\eqref{eq:la}, both quantities are a $\compRatio_n$ fraction of $a_n$, which in turn is at most $M_n$, the optimal  (effective) efficiency.
%
\end{proof}

\begin{figure}[htbp] %
\text{\hspace{-0.30in}}
\includegraphics[height=3in]{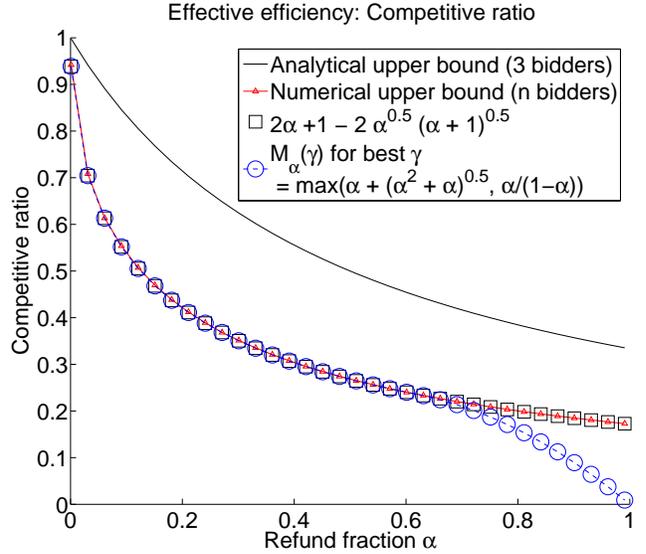}
\caption{Effective efficiency (EE) bounds as a function of $\alpha$. 
The top curve is $\compRatio_3 = 1/(1+2\alpha)$. 
The middle curve is a numerical upper bound on any deterministic algorithm's EE: it shows $\compRatio_n$ computed numerically from its recursion relations. 
The bottom curve shows (a {\em lower} bound on)  our algorithm's EE for the best $\gamma_\alpha$: it matches the upper bound for $\alpha < 0.618$. 
Our choice of  $\gamma$ is constrained by $\alpha < \gamma/(\gamma+1)$; if it were not, the bounds would match for all $\alpha$.
}
\label{fig:effEff}
\end{figure}

Figure~\ref{fig:effEff} strongly suggests that the competitive ratio of any algorithm cannot be higher than $2\alpha +1 - 2 \alpha^{0.5} (\alpha + 1)^{0.5}$, shown as squares in the figure. Note that for this $\compRatio$  the characteristic equation of Eq.~\eqref{eq:rec} has a double root.

The triangles plot the minimum $\compRatio$ found for the corresponding $\alpha$ for different values of $n$ (we used Fibonacci values up to rank 12, i.e. largest $n$ was 144). The $\compRatio$ values were found via binary search. It was true in general, although not always, that the higher $n$, the lower $\compRatio_n$. 
We suspect that 
one can always find an increasing sequence of integers $\{n_i\}_{i \geq 1}$ such that a solution $\compRatio_{n_i}$ to Eqs.~\eqref{eq:rec} and~\eqref{eq:la} converges from above to $2\alpha +1 - 2 \alpha^{0.5} (\alpha + 1)^{0.5}$ as $i \to \infty$.

Lemma~\ref{result:w(S)} implies that for our algorithm
\[
u =  w(S) - \alpha w(R) 
\geq w(S) - \alpha w(S)/\gamma 
\geq \frac{\OPT/}{1+\gamma} \left(1-\frac{\alpha}{\gamma} \right).
\]
\newcommand{\lobo}[1]{\ensuremath{ \underline{u}(\gamma_{#1})}}
Let $\lobo{} =  \frac{1}{1+\gamma} \left(1-\frac{\alpha}{\gamma} \right)$. 
Subject to the constraint $\alpha \leq  \frac{\gamma}{\gamma + 1}$, $\lobo{}$ is maximized for $\gamma_0 = \max\{\alpha + \sqrt{\alpha^2 + \alpha}, \frac{\alpha}{1-\alpha}\}$. 
$\lobo{0}$ is displayed in Fig.~\ref{fig:effEff} by circles. 
The value 0.618 (the golden ratio) is where $\frac{\alpha}{1-\alpha}$ becomes higher than $\alpha + \sqrt{\alpha^2 + \alpha}$. 
If $\alpha < 0.618$, $\lobo{0} = 2\alpha +1 - 2 \alpha^{0.5} (\alpha + 1)^{0.5}$, which matches the numerical upper bound. 
Recall that this is just a worst-case lower bound on the effective efficiency, but likewise, so are the upper bounds.
The top curve plots $\compRatio_3 = 1/(1+2\alpha)$.

Recall that when $\alpha=0$ all bidders  can be tentatively accepted (by letting $\gamma=0$) since they incur no loss and do not have to be refunded. Then, the optimal matching can be found via a one-shot (offline) algorithm at time $T$. 




\section{Revenue of \gx/}
\label{sec:rev}

We show that apart from favorable incentive and efficiency
properties,  \gx/ is also competitive with respect to revenue.

As a revenue benchmark, we consider the offline VCG mechanism because it 
generates the highest revenue among truthful efficient allocation mechanisms~\cite{krishna02auction}.
We show that our mechanism is competitive with respect to revenue with VCG on bidders' true {\em values}. 

\newcommand{\aVec}{\biwo'}
Let $\aVec$ be a sequence of bids - when defining VCG on $\aVec$ we will assume that all bids are received at once by VCG. 
Let $\aVec_{-i}$ denote
the set of all bids in $\aVec$ except bidder $i$'s.
The VCG mechanism implements an efficient allocation and thus the matching it outputs is optimal.
If $i \in \OPT/[\aVec]$ then VCG charges bidder $i$ its  externality on the other bidders: 
\begin{equation}
\label{eq:VCG}
\sum_{k \in \OPT/[\aVec_{-i}]} \aVec(k)
\quad - 
\sum_{j \neq i, j \in \OPT/[\aVec]} \aVec(j) 
\end{equation}

We will use the following known (see e.g.~\cite{flaxman}, Fact 3.2) combinatorial property of our setting: $\all i\neq x$, if $x\in \OPT/[\aVec]$ then $x\in \OPT/[\aVec_{-i}]$.

\begin{lemma}
\label{lemma:VCGMatroid}
A winning bidder's VCG payment is a losing bid. Also, the VCG revenue can only increase if some bids in $\aVec$ are increased.
\end{lemma}

On bids $\aVec$ we denote the VCG revenue  by $\rvcg[\aVec]$ and the net revenue of \gx/ (payments from survivors minus bump payments) by $\rgi[\aVec]$. 
%
\begin{theorem}
\label{thm:rev}
Assume $w(i) \geq v(i) \all i$, i.e. no one bids below their true value, since that would be dominated.
Then
\[
\rgi[\biwo] \geq \frac{1-\alpha - \alpha/\gamma}{1+\gamma} \rvcg[\biwo]
\]
\end{theorem}

This theorem shows the tradeoff between $\gamma$, the improvement factor required for bumping an accepted bidder and $\alpha$, the fraction returned
as the bump payment. For instance, for $\alpha=0.25$ (refund of 1/4th the bid), if we choose $\gamma=1$ then the constant in Theorem~\ref{thm:rev} becomes $0.25$, i.e. our mechanism  obtains at least a quarter of the VCG revenue. 

We will now prove Theorem~\ref{thm:rev}. 

Lemma~\ref{result:w(S)} implies that payments received by $\gx/$ are at least
$\SV(1 - \alpha)$ (since only survivors pay) and that bump payments sum to at most $\SV{\displaystyle \frac{\alpha}{\gamma}}$. 
It will suffice to show that $\SV  \geq \rvcg[\biwo] / (1+\gamma)$. 



\newcommand{\someVec}{\mathbf{u}}
\newcommand{\hsp}{\hspace{-0.09in}}
Let
 $\someVec(i)=
\begin{cases}
\max(\sv{i}, w(i)/(1+\gamma)), & \mbox{\hsp if $i\in S$} \\
\frac{w(i)}{1+\gamma}, & \mbox{\hsp if $i\notin S$} 
\end{cases}.
$

 Lemma~\ref{lemma:funnyvcg} states 
that $\all s \in S$, $s \in \OPT/[\tilwo{\biwo}]$. Since on $\tilwo{\biwo}$ VCG payments cannot be higher than its efficiency, $\SV \geq \rvcg[\tilwo{\biwo}]$. Also, 
Lemma~\ref{lemma:VCGMatroid} implies
\[
\rvcg[\tilwo{\biwo}] 
= 
\rvcg[\someVec] 
\geq  
\rvcg[\biwo]/(1+\gamma),
\]
%
since when going from $\tilwo{\biwo}$ to $\someVec$ only VCG winners may increase their bid, and 
for all $i$, $\someVec(i) \geq w(i)/(1+\gamma)$.

Note that unlike the analogous efficiency result (Theorem~\ref{prop:effTruePos}), this result makes no 
  assumption on bidders' utilities.
\section{Speculators}
\label{sec:spec}

Since money  is given away, speculators, that is, bidders without interest in any item are likely to enter the mechanism looking for bump payments. 
For a speculator $i$, utility is also given by Eq.~\eqref{eq:ut}, but with $v(i)=0$.  
Speculators may bid (under false identities) more than once or collude.
Their bids can effectively induce reserve prices, since actual bidders will have to bid a $1+\gamma$ factor higher than a competing speculator. 
 If speculators bid judiciously on high-demand items, 
they can garner payment from the auctioneer, who  gets even more revenue via
larger prices for high-value bidders. So, it is clear that speculators
are impactful. 

In this section we address the question of how speculators affect the
mechanism. In Section~\ref{sec:specpos}, we show two positive results
\begin{itemize}
\item
We show that the \gx/ algorithm has good overall efficiency, as long as the
speculators have positive overall surplus and the survivors are
best-responding.
\item
We prove a bound on the overall revenue that speculators can obtain. 
\end{itemize}
In Section~\ref{sec:specneg} we give a more detailed discussion of speculator strategy.  Along the way,  
we show that many natural simplifying assumptions regarding
speculators' or bidders' strategies are unfortunately false. 
Specifically, we show that:
\begin{itemize}
\item the profits available for speculators may depend on the arrival order of actual bidders (Example~\ref{ex:C11C});
\item there may be no pure Nash equilibrium for actual bidders or speculators (Example~\ref{ex:C11C});
\item speculators may prefer to induce a suboptimal perfect matching of actual bidders (Example~\ref{ex:suboptGoodSpec});
\item a colluding set of speculators may be able to get higher bump payments if some of them 
survive (Example~\ref{ex:sacr}).
\end{itemize}

\subsection{Impact on Efficiency and Revenue}
\label{sec:specpos}
 
The following result gives a competitive ratio of our algorithm's efficiency with respect to the optimum efficiency given bidders' {\em true} values.
It only requires that total speculator utility is non-negative: this is particularly applicable if speculators are coordinated and can make monetary transfers between them.
\begin{proposition}
\label{prop:effTrueSpecPosSurvTru}
Let $\biwo$ be a set of bids such that actual survivors are best-responding and total speculator surplus, i.e. the sum of speculators' payments minus the sum of speculators' prices, is non-negative. 
Then the \gx/ algorithm with {\em true values} $\trwo$ has efficiency
\[
\sum_{i \ \mathrm{ actual} ,  i \in S} v(i) \geq \frac{1 - \alpha - \frac{\alpha}{\gamma} }{(1-\alpha)(1+\gamma)} \cdot \OPT/[\trwo]
\]
\end{proposition}
The proof is mostly algebraic and deferred to the Appendix.  
This result is a strengthening of Theorem~\ref{prop:effTruePos}: 
Prop.~\ref{prop:effTrueSpecPosSurvTru}'s constant is larger and its preconditions are less general.
 Note that Prop.~\ref{prop:effTrueSpecPosSurvTru} requires that {\em actual} survivors are best-responding; a speculator's best response cannot induce it to survive.

Next we prove an upper bound on speculators' profit:
\begin{proposition}
\label{prop:aOPTg}
Speculators' total profit is at most  $\alpha \OPT//\gamma$.
\end{proposition}
\begin{proof}

Let $\Sigma$ be the sum of survival weights for speculators that have survived. Denote  speculators' profit by $\Pi \leq -(1-\alpha)\Sigma+\alpha w(R)$, where $R$ are the participants who obtain bump payments (some may be true bidders). 

By Lemma~\ref{result:w(S)}, $w(R) \leq (\Sigma + A)/\gamma$,
where $A$ is the total weight of survivors that are actual bidders. 
We get 
\[
\Pi \leq -(1-\alpha- \frac{\alpha}{\gamma})\Sigma + \frac{\alpha}{\gamma} A
\]
The claim follows since $ (1-\alpha- \frac{\alpha}{\gamma})\Sigma \geq 0$ and $A \leq \OPT/$.
\end{proof}

\subsection{Speculator Strategies}
\label{sec:specneg}

At first glance, it would seem that it is in the speculators' best
interest to induce an assignment of actual bidders of weight as high as
possible in the survivor set, since then overall bump payments would be maximized.
This is true in some cases
but not always (Example~\ref{ex:suboptGoodSpec}). The reason for such a distinction is that the {\em order} of bidders arriving also influences the maximum refunds attainable by speculators as shown below.
%
\begin{example}
\label{ex:C11C}
Consider two bidders, one bidding $1$, the other $C>1$, on two items and assume that  speculators cannot collude. 
If $C$ arrives first, 
no speculator can have higher revenue if bumped than when bidding $1/(1+\gamma)$ on both items: this is actually a Nash equilibrium (NE) for them.
If $1$ however arrives first, then speculators could participate with two identities bidding $1/(1+\gamma)$ and $C/(1+\gamma)$ on both items, both  being bumped. 
One can show via a case analysis that there is no pure strategy NE for speculators. 
\end{example}
This example also shows that there may not be a pure strategy NE when only actual bidders participate: if two bidders with low values arrive, followed by the 1 bidder and after that the $C$ bidder, then the two low value bidders are essentially speculators and the argument in the example applies.

Observe that
a speculator who is bumped with a bid of $x$ could have obtained more bump payment by entering an earlier bid of at most $x/(1+\gamma)$; likewise, he could have obtained yet more by bidding earlier $x/(1+\gamma)^2$; and so on. This is formalized as follows.
\begin{definition}
Let $x>0$. We say that the speculator $\sigma$ is an {\em $x$-geometric speculator} with choice set $N(i)$ if $\sigma$ places bids as follows on choice set  $N(i)$. 
Let $\eps$ be the minimum strictly positive bid that can be made and 
\[
l = 1+\left\lfloor \frac{\log(x/\eps)}{\log(1+\gamma)} \right\rfloor
\mbox{\ i.e. \ $l\in Z$ }  \& \\
\frac{x}{(1+\gamma)^{l}} \geq \eps >
\frac{x}{(1+\gamma)^{l+1}}
\]
Then $\sigma$ participates with $l+1$ different identities, placing consecutive bids of 
$\frac{x}{(1+\gamma)^{l}},
\frac{x}{(1+\gamma)^{l-1}}, \dots, 
\frac{x}{(1+\gamma)},
x$
on $N(i)$. 
\end{definition}

 If speculators have full information on bidders' values and bidders in \OPT/ arrive in increasing order of their values,
the outcome has many desirable properties: 
\begin{lemma} 
\label{prop:OPTincr}
Fix a set of actual bids such that  \OPT/$[\trwo]$ bids arrive in increasing order.
%
Suppose that speculators collude and want to maximize their joint revenue. Then optimal speculator bidding has the following consequences:
\begin{itemize}
\item no speculator survives, no actual bidder is bumped; all \OPT/ bidders and only them are accepted.
\item speculators can achieve the highest payoff possible as given by Lemma~\ref{prop:aOPTg}.
\item truthful bidding is a NE for all actual bidders.
\end{itemize}
\end{lemma}
This is a further strengthening of Theorem~\ref{prop:effTruePos}. 
Optimal speculator bidding in this case is as follows.
 For each bidder $i\in \OPT/$ with choice set $N(i)$ there will be one $w(i)/(1+\gamma)$-geometric speculator $\sigma_i$ with the same choice set. The proof is deferred to the full version. 

This result has an appealing interpretation. If very well informed, speculators can overcome the efficiency loss due to late bidders not being able to improve by a $1+\gamma$ factor over their earlier competitors.

In general however, speculators may prefer to induce a suboptimal perfect matching: 
%
%
\begin{example}
\label{ex:suboptGoodSpec}
Consider two items $\{i_1, i_2\}$ and three bidders $b_1, b_2, b_3$ arriving in this order; bidder $k$ is interested in item $i_k, k=1,2$, while bidder $3$ is interested in any of $i_1$ or $i_2$.
Note that any matching that does not match all three bidders is valid. 
Assume that  $w(b_1) < w(b_3) < (1+\gamma) w(b_1)$ and $w(b_2) > 2 w(b_3)$. 
The following analysis shows that speculators prefer the suboptimal set of actual bidders $b_1$ and $b_2$ to the optimal one with $b_2$ and $b_3$.
\begin{itemize}
\item If both $b_2$ and $b_3$ survive, then speculators' profit is at most $2w(b_3)/\gamma$: the speculator bumped by $b_2$ must have a lower weight than the one bumped by $b_3$, which is at most $w(b_3)/(1+\gamma)$. Even if speculators are geometric,  speculator profit can only go as high as $2w(b_3)/\gamma$.
\item If however $b_1$ and $b_2$ are alive when $b_3$ arrives, $b_3$ cannot bump $b_1$. By simply having one geometric   $w(b_2)/(1+\gamma)$-speculator  which is bumped by $b_2$, speculator profit is $w(b_2)/\gamma > 2w(b_3)/\gamma$.
\end{itemize}
\end{example}

The following example shows that speculators may be able to make more money if they ``sacrifice'', i.e. some of them intentionally survive so that others obtain high refunds:
%
%
\begin{example} 
\label{ex:sacr}
Let there be $k$ items, $k-1$ actual bidders bidding $C>1$ all arriving before an actual bidder bidding $1$; all $k$ bidders bid on all the items. If speculators coordinate and participate with $k$ identities as $C/(1+\gamma)$-geometric speculators on all the items then  total speculator payoff is
\[
(k-1) \alpha C/\gamma - (1-\alpha) C/\gamma = (k\alpha - 1) C/\gamma
\]
since $k-1$ will be bumped, but one will survive. 
If no speculator survives, the most money speculators can make is $k/\gamma$, by participating as $k$ $1/(1+\gamma)$-geometric speculators. For any $\alpha > 1/k$,  for a large enough $C$, speculators' profit is higher when one of them is sacrificed. 
\end{example}

\section{Other Game-Theoretic Claims}

We will now show that several appealing statements regarding incentives in our algorithm  are false.

The algorithm may be more appealing for incentive purposes if we paid a bumped bidder $\alpha \sv{i}$ instead of $\alpha w(i)$ as bump payment. The following example shows why this may result in a deficit: 
\begin{example}
Fix $\alpha$ and consider an early bidder $e$ bidding $1$ and a late
bidder $\ell$ bidding $L$ on one item where $L >
(1+\gamma)^2/\alpha$. Bidder $\ell$ survives and pays $(1+\gamma)$. If we were
to refund $e$ an $\alpha$ fraction of $\sv{e}$, $e$ would get $\alpha
L/(1+\gamma)$. The choice of $L$ ensures that $e$ is paid more than
$\ell$ pays, i.e. the mechanism runs a deficit.
\end{example}

We assumed throughout that as soon as a bidder arrives, its choice set is known. If however that is private information as well, incentives become weaker: 
in Example~\ref{ex:32}, no bid by $\sta{B}$ on its true item $i_2$ is a best-response if bidding on different item(s) instead is allowed. This example also suggests why a naive generalization of \gx/ to the setting where bidders have a different value for each of several items would not be able to incentivize bidders to bid at least their true value for each item.
%
%
\begin{example} 
\label{ex:32}
\newcommand{\boh}{\ensuremath{B^{-3/2}}} 
 
Consider two items $i_1, i_2$ and the following set of three bidders (arriving in this order): \boh with value $(1+\gamma)^{-3/2}$ for any of ${i_1, i_2}$ (only demanding one of them), $\sta{B}$ who has value $x < \alpha (1+\gamma)^{-3/2}$ for item $i_2$ 
 and $B^1$ bidding 1 on item $i_1$. Assume \boh and $B^1$ bid truthfully. We will show that, whenever $\sta{B}$ bids on $\{i_1, i_2\}$, it can do {\em strictly} better by bidding on $i_1$ only.

We claim that if $\sta{B}$  bids on $\{i_1, i_2\}$ 
then its utility is at most $\alpha (1+\gamma)^{-3/2}$. This is clear if it survives. If it is bumped by $B^1$, then its bid cannot be higher than $(1+\gamma)^{-3/2}$ (\boh's bid), since $B^1$ can replace any of $\boh$ and $\sta{B}$. But then $\sta{B}$'s compensation is at most $\alpha(1+\gamma)^{-3/2}$. 
Let $0 < \eps < 1/2$. By bidding $(1+\gamma)^{-1-\eps}$ on $i_1$ only and being bumped by $B^1$, $\sta{B}$ can get utility $\alpha (1+\gamma)^{-1-\eps} > \alpha (1+\gamma)^{-3/2}$. 
\end{example}

We have however the following conjecture: if a bidder prefers surviving to being refunded, 
they are better off bidding on their true choice set.

If bidders myopically and simultaneously best-respond then these dynamics may lead to bid vectors where the sum of their utilities is negative:
%
%
\begin{example}
\label{ex:myopBR}
Let there be $n$ items and $2n$ bidders (interested in any item) arriving in increasing order of their values: 
bidders $B_1, \dots,  B_n$ have value $x < \min\{1, 5(1-\alpha) \}/(1+\gamma)$, bidders $B_{n+1}, \dots, B_{2n-1}$ 
have value $1$ and bidder $B_{2n}$ has value $5$. Assume they all bid truthfully initially.

Bidders $n+1, \dots,2n$ are best-responding by Lemma~\ref{prop:BR} and they will not change their bids.
Each bidder $B_i$'s, $i\in [1,n]$, myopic best-response is to bid $5/(1+\gamma)$.  Only one of them will be bumped by 
$B_{2n}$ and the others will survive: the sum of their utilities is at most 
$x - (n-1) \cdot 5(1-\alpha)/(1+\gamma) < -(n-2) \cdot 5(1-\alpha)/(1+\gamma)$. No bidder between $n+1$ and $2n-1$ will be 
accepted. Since $B_{2n}$'s utility is at most $5$, for large  $n$ the sum of all $2n$ bidders' utilities will be negative.
%
%
\end{example}
This example does not preclude good performance for other dynamics.



\section{Concluding Remarks}
Advertisers seek a mechanism to reserve ad slots in advance, while the publishers present a large 
inventory of ad slots with varying characteristics and seek automatic, online methods for 
pricing and allocation of reservations. In this paper, we present a simple model for auctioning such ad slots in advance,
which allows canceling allocations at the cost of a bump payment. We present an efficiently implementable online mechanism to derive prices and 
bump payments that has many desirable properties of
incentives, revenue and efficiency. 
These properties hold even though we may have speculators who are in the game for earning bump payments only. 
Our results make no assumptions about 
order of arrival of bids or the value distribution of bidders. 

Our work leaves open
several technical and modeling directions to study in the future.
From a technical point of view, the main questions are about designing
mechanisms with improved revenue and efficiency, perhaps under 
additional assumptions
about value distributions and bid arrivals. Also, mechanisms that limit 
further the role of speculators will be of interest. In addition, 
there are other models that may be applicable as well. Interesting directions for future research include  allowing bidders to pay more for higher $\gamma$ (making it harder for future bidders to displace this bidder) or higher $\alpha$ (being refunded more in case of being bumped). Other mechanisms may allow  $\alpha$ to be a function of time between the acceptance and 
bumping. Accepted advertisers may be allowed to withdraw their bid
at any time. There may be a secondary market where bidders may buy insurance against cancellations.
Finally, advertisers may want  a bundle of slots, say many impressions at multiple websites simultaneously, which will result in combinatorial
extension of the auctions we study here. We believe that there is a rich 
collection of such mechanism design and analysis issues of interest
which will need to inform any online system for advanced ad slotting
with cancellations.

\textbf{Acknowledgments.} We would like to thank Stanislav Angelov for pointing us to~\cite{McGregor} and the EconCS research group at Harvard for helpful discussions.

\bibliographystyle{abbrv}
\bibliography{refund5}

\appendix

\medskip

\setcounter{section}{1}

\subsection{Proof of Prop.~\ref{prop:effTrueSpecPosSurvTru}}

Assume that actual survivors bid honestly: $w(i) = v(i) \all i$; at the end of the proof we will eliminate this assumption.

Let $w'(i)=
\begin{cases}
\sv{i}, & \mbox{ if $i\in S$ and $i$ is a speculator} \\
w(i), & \mbox{ otherwise} 
\end{cases}
$. 

As $\sv{i}$ can only be $(1+\gamma) w(k)$ if $i$ bumps $k$ or $w(i')$ if $i'$ is bumped instead of $i$, the set of survivors will still be $S$ if the mechanism is run on $\biwo'$ instead of $\biwo$. We have
\begin{equation}
\label{eq:spsv_tr}
w'(S) = {\textstyle \sum_{i \in S} w'(i)} \geq  \frac{\OPT/[\biwo']}{1+\gamma} \geq  \frac{\OPT/[\trwo]}{1+\gamma} 
\end{equation}
The first inequality follows from Lemma~\ref{result:w(S)}. Each actual bidder $i$ bids at least its true value: $w'(i) \geq v(i)$  and there is the additional competition of speculators; this fact yields the second inequality. Also,
\begin{align*}
 \gamma w(R) \leq \SV &\leq \sum_{\substack{i \ \mathrm{ actual} \\ i \in S}} w(i) + \sum_{\substack{i \ \mathrm{ speculator} \\ i \in S}} \sv{i} = w'(S)
\end{align*}
Again, the first inequality follows from Lemma~\ref{result:w(S)}.
By assumption, speculator payments (a $(1-\alpha)$ fraction of their survival weights)
 cannot be higher than speculator refunds:
\begin{align*}
 (1-\alpha) \sum_{\substack{i \ \mathrm{ spec.} \\ i \in S}} \sv{i}
\leq 
\sum _{\substack{i \ \mathrm{ spec.} \\ i \in R}} \alpha w(i) 
\leq 
\sum _{i \in R} \alpha w(i) = \alpha w(R)
\end{align*}
By combining the last two relationships, we get
\[
 (1-\alpha) \sum_{i \ \mathrm{spec.} , i \in S} \sv{i} 
\leq
\frac{\alpha}{\gamma}
 w'(S)
\]
Adding $(1-\alpha) \sum_{i \ \mathrm{ actual} , i \in S}  w(i) $ to both sides we get
\begin{align*}
(1-\alpha) w'(S) 
&\leq
 \frac{\alpha}{\gamma} w'(S) + (1-\alpha) \sum_{i \ \mathrm{actual} ,i \in S}  w(i) \quad \mbox{i.e. } \\
\frac{1 - \alpha - \frac{\alpha}{\gamma} }{1-\alpha}w'(S) 
&\leq
\sum_{i  \ \mathrm{actual} , i \in S}  w(i) = 
\sum_{i  \ \mathrm{actual} , i \in S}  v(i)
\end{align*}
The last equality follows since actual survivors are bidding truthfully. This last inequality, together with Eq.~\eqref{eq:spsv_tr} imply the proposition's claim.

Lemma~\ref{prop:BR} shows that bidding truthfully is a (weak) best-response for survivors. Let $i$ be an actual survivor. Any bid above its true value is also a best-response for $i$. In the claim, as we increase $i$'s bid, 
the right-hand side quantity increases (if at all) at a constant rate which is less than 1, the left-hand side quantity's increase rate.
%

\subsection{Proof of Lemma~\ref{lemma:funnyvcg}}

\newcommand{\svt}[2]{\ensuremath{w^{\mathrm{sv}}_{{\scriptscriptstyle \leq} #2}(#1)}} 

We will denote by $\svt{b}{t}$ the minimum bid bidder $b$ must make in order to survive up to and including time $t$. Then $\ac{b} = \svt{b}{b}$ and $\sv{b} = \svt{b}{T}$. It is clear that $ \svt{b}{t} \leq  \svt{b}{t+1}$.

\begin{definition}
Let $B$ be a set of bidders.
We say that $B$ is {\em tight} for a bidder $i$ at time $t$ if all bidders in $B$ are alive at $t$, $B$ can be matched but $B\cup\{i\}$ cannot be matched. 
We say that $B$ {\em $\tilwo{\biwo}$-dominates a bidder $i$ at time $t$} in the algorithm if $B$ is tight for $i$ at $t$ and  $\all b\in B$, 
$\svt{b}{t} \geq w(i)/(1+\gamma)$.
\end{definition}

\begin{lemma}
\label{lemma:RtTight}
 $X_t$ is tight for $t$ at $t$.
\end{lemma}
\begin{proof}
$X_t$ can be matched since $X_t\subseteq A_{t-1}$.

Suppose for a contradiction that $X_t \wi{t}$ can be matched. Then $X_t \neq A_{t-1}$ since $A_{t-1}$ is a perfect matching by assumption. 
Therefore there exists $X \subset A_{t-1} \setminus X_t, |X| = |A_{t-1}| - |X_t| -1$ such that $X_t \wi{t} \cup X$ can be matched. There exists exactly one bidder $\{y\} =  A_{t-1} \setminus (X_t \cup X)$ and we have that  $X_t \wi{t} \cup X = A_{t-1} \wi{t} \wo{y}$ is a perfect matching, implying $y \in X_t$, contradiction.
\end{proof}

Let $\sta{i}$ be the time step when $i$ ceases to be alive (i.e. $i$ if $i$ is not accepted or the time $i$ is bumped if $i$ was accepted). We inductively construct a sequence $\{B_t\}_{\sta{i} \leq t\leq n}$ as follows: if $i$ is not accepted, $B_{i} = X_{i}$; if $i$ is bumped by $\sta{i}$ then $B_{\sta{i}} = X_{\sta{i}}\wi{\sta{i}}\wo{i}$. At time $t \geq \sta{i}+1$,
\begin{itemize}
\item if no bidder in $B_{t-1}$ is bumped, then we let $B_t = B_{t-1}$.
\item if $t$ bumps some $b\in B_{t-1}$ then we let $B_t = (B_{t-1} \cup X_t \wi{t}) \wo{b}$
\end{itemize}

We will prove inductively on $t$ that 
\begin{lemma}
\label{lemma:t_block}
$B_t$ \tilwo{\biwo}-dominates $i$ at time $t$.
\end{lemma}
\begin{proof}[of Lemma~\ref{lemma:t_block}]
Note that by definition, all bidders in $B_t$ are alive at $t$.

\noindent \textbf{Base case $t=\sta{i}$:} 


If $i$ is not accepted ($\sta{i}=i$), $i$ cannot bump any bidder in $X_i$: therefore $\all b\in X_i, \svt{b}{i} \geq w(i)/(1+\gamma)$. $X_{i}$ is tight for $i$ at $i$ by Lemma~\ref{lemma:RtTight}. 

If $i$ is bumped, then $w(i) \leq \svt{r}{\sta{i}}, \all r \in X_{\sta{i}}$. $B_{\sta{i}} = X_{\sta{i}}\wi{\sta{i}}\wo{i}$ can be matched since they are all alive at $\sta{i}$. $X_{\sta{i}}\wi{\sta{i}}$ cannot be matched since otherwise $\sta{i}$ would not need to bump $i \in X_{\sta{i}}$.

\noindent \textbf{Inductive step:}  Assume that $B_{t-1}$ \tilwo{\biwo}-dominates $i$ at $t-1$. If at time $t$, no bidder in $B_{t-1}$ is bumped, then the claim obviously holds by the induction hypothesis.
Otherwise, let $b\in B_{t-1}$ be the bidder that is bumped by $t$. 

Clearly, $(B_{t-1} \cup X_t \wi{t}) \wo{b}$ can be matched since they are alive at $t$. Suppose for a contradiction that $B_t \wi{i}= (B_{t-1} \cup X_t \wi{t}) \wi{i} \wo{b}$ could be matched. $i \notin B_t$ since $i$ is no longer alive. $B_{t-1} \cup X_t$ can be matched since they are all alive at $t-1$. As $|B_{t-1} \cup X_t| = |B_t \wi{i}| - 1$, either $B_{t-1} \cup X_t \wi{i}$ or $B_{t-1} \cup X_t \wi{t}$ can be matched. The first case is not possible since a subset, $B_{t-1} \wi{i}$, cannot be matched (by the induction hypothesis); the second case is not possible since $X_t \wi{t}$ cannot be matched  (Lemma~\ref{lemma:RtTight}). We have reached a contradiction, so $B_t$ must be tight for $i$.

By the induction hypothesis, $\all b' \in B_{t-1},$ $\sv{b'}_{\leq t-1} \geq w(i)/(1+\gamma)$. 
As noted before, survival thresholds can only increase from $t-1$ to $t$ 
and $w(t) \geq (1+\gamma) w(b)$. \qed
\end{proof}

We are ready for 
\begin{proof}[of Lemma~\ref{lemma:funnyvcg}]

Let $V$ be the \OPT/[\tilwo{\biwo}] assignment (where ties are broken in favor of bidders in $S$).
%
Suppose for a contradiction that there exists a non-survivor $i \in V$. By Lemma~\ref{lemma:t_block} for time $n$, $i$ is dominated by  a set $B_n \subseteq S$ at time $n$. 
Since $i\notin S$, but $B_n \subseteq S$, in $\tilwo{\biwo}$ any bidder in $B_n$ has a higher weight than $i$.

Since $V$ is a perfect matching and $B_n$ can be matched there must exist $V' \subset V\setminus B_n, |V'| = |V|-|B_n|$ ($V' = \emptyset$ if $B_n$ is a perfect matching) such that $B_n \cup V'$ is a (perfect) matching. We know that $B_n \wi{i}$ cannot be matched, therefore $i \notin V'$. However, $i \in V$ therefore $i \in V\setminus V'$. $V\wo{i}$ can be matched and has size $|V|-1$. Therefore there $\exists b\in B_n \cup V'$, $b \notin V \wo{i}$ such that  $V\wi{b} \wo{i}$ can be matched. That implies $b \in B_n \subseteq S$, i.e. $\tilwo{\biwo}(b) \geq \tilwo{\biwo}(i)$.  But then $V\wi{b} \wo{i}$ is a perfect matching of higher weight than $V$, contradiction.

That is, $V\setminus S = \emptyset$, i.e. $V=S$ since both are perfect matchings.
\end{proof}

\end{document}